\documentclass[conference,a4paper]{APSIPA2021}
\usepackage{amsmath}
\usepackage{graphicx}
\usepackage{multirow}
\usepackage{booktabs} 
\usepackage{threeparttable}
\usepackage[backend=biber,style=ieee,]{biblatex}

\usepackage{array}

\usepackage{geometry}
\usepackage{fancyhdr}

\fancypagestyle{firststyle}{
  \fancyhf{}
  \fancyhead[C]{2024 Asia Pacific Signal and Information Processing Association Annual Summit and Conference (APSIPA ASC)}
}

\begin{document}

\title{Two-stage Framework for Robust Speech Emotion Recognition Using Target Speaker Extraction in Human Speech Noise Conditions}

\author{
\authorblockN{
Jinyi Mi, Xiaohan Shi, Ding Ma, Jiajun He, Takuya Fujimura and Tomoki Toda
}

\authorblockA{
% \authorrefmark{1}
Nagoya University, Japan \\
E-mail: \{mi.jinyi, xiaohan.shi, ding.ma, jiajun.he, fujimura.takuya\}@g.sp.m.is.nagoya-u.ac.jp, tomoki@icts.nagoya-u.ac.jp}

% \authorblockA{
% \authorrefmark{2}
% University of Macau, Macau \\
% E-mail: yy@um.edu.mo  Tel/Fax: +853-XXXXXXXX}
}

\maketitle
\thispagestyle{firststyle}
\pagestyle{fancy}

\begin{abstract}
Developing a robust speech emotion recognition (SER) system in noisy conditions faces challenges posed by different noise properties. Most previous studies have not considered the impact of human speech noise, thus limiting the application scope of SER. In this paper, we propose a novel two-stage framework for the problem by cascading target speaker extraction (TSE) method and SER. We first train a TSE model to extract the speech of target speaker from a mixture. Then, in the second stage, we utilize the extracted speech for SER training. Additionally, we explore a joint training of TSE and SER models in the second stage. Our developed system achieves a 14.33\% improvement in unweighted accuracy (UA) compared to a baseline without using TSE method, demonstrating the effectiveness of our framework in mitigating the impact of human speech noise. Moreover, we conduct experiments considering speaker gender, showing that our framework performs particularly well in different-gender mixture.
\end{abstract}

\section{Introduction}
\label{sec:intro}
Speech is a significant part of human communication. Besides linguistic information, it contains unique paralinguistic information such as gender, emotion, and age, which is essential to the normal communication. In certain instances, misunderstanding paralinguistic features would distort the correct information conveyed by speech, leading to an ineffective communication. Therefore, it is necessary to develop human-like communication machines that can comprehend paralinguistic data.

Speech emotion recognition (SER), as a branch of affective computing, has garnered growing attention over the past two decades because of its contribution to human-computer interactions \cite{Schuller18_Review, shi2020dimensional}. Generally, the mechanism of SER involves extracting and classifying effective emotional features from audio signals so that various emotions of a speaker can be captured, thanks to which SER has been applied in healthcare \cite{Uddin20_healthcare, Hossain15_healthcare}, driver safety \cite{Grimm07_driver, Tawari10_driver}, call center \cite{Petrushin99_call, Morrison07_call}, and online education \cite{Li07_education, Tickle13_education}. At present, research on SER systems on scenarios devoid of background noises, often referred to as clean scenarios, has shown good performance \cite{Liu23_Discriminative, Shen23_Temporalshift, Chen23_Exploring}. However, in real-world environments, the performance of SER significantly degrades, mainly due to the presence of various noises from different sources. These unknown noises severely affect the performance of SER systems, which poses major challenges for the widespread application of SER systems.

Several studies have focused on SER tasks in the environment affected by specific noise sources, including communication systems \cite{Huang13_WhiteNoise, Liu23_FactoryNoise}, transportation \cite{Chenchah17_TransportationNoise, Liu23_FactoryNoise}, and industrial activities \cite{Liu23_FactoryNoise}. In \cite{Huang13_WhiteNoise}, Huang et al. studied SER from speech signals with additive white Gaussian noise, they proposed two speech enhancement methods based on spectral subtraction and masking properties, respectively. In \cite{Chenchah17_TransportationNoise}, Chenchah et al. used power-normalized cepstral coefficients as acoustic features for improving the robustness of SER systems in noisy environment from cars and trains. In \cite{Liu23_FactoryNoise}, Liu et al. proposed a multi-level knowledge distillation framework, which significantly reduced the affects of noises from channel, car, and factory.

However, the aforementioned studies mainly concentrate on addressing the noise sources associated with non-human activities, leaving a gap in addressing prevalent sources of noise in human-centric environments. While Shi et al. \cite{Shi23_ImpulsiveNoise} did adopt ASR representations to filter out a specific category of noise related to human activities, which is typically from human physical actions like knocking on doors, a more common category of noise stemming from human speech itself remains underexplored. This type of noise called human speech noise, which is common in activities involving human interactions such as social gatherings, often becomes entangled with target speech data, forming a complex acoustic environment. Therefore, human speech noise becomes more unpredictable and more challenging to address. 

On the other hand, humans have an extraordinary ability to selectively concentrate on a single speaker among a complex acoustic environment, commonly called cocktail party effect \cite{Handel93_Cocktail}. To replicate this specialized listening ability in machines, target speaker extraction (TSE) technique has been developed. This technique exploits an auxiliary information of the target speaker and extracts speech of that speaker from the mixture. \v{S}vec et al. \cite{vsvec2022analysis} explored the potential of TSE for extracting target emotional speech. In light of this, we propose a novel two-stage framework by cascading TSE method and SER to mitigate the impact of human speech noise. In the experiments, we utilize the TSE--SER framework to ShiftCNN \cite{Shen23_Temporalshift} that is a state-of-the-art SER model, and compare its performance to verify the effectiveness of our framework. Furthermore, we investigate different factors on the performance of SER, including training methods and speaker genders. Our contributions to this work are as follows:

\begin{itemize}
\item We apply the TSE method on SER tasks to address the practical scenario in which target speech is interfered by human speech noises. To our knowledge, this study is the pioneering effort in exploring the integration of TSE with SER.
\item We propose a two-stage framework using different training methods. According to comparative experiments against the baselines without using TSE technique, our framework significantly improves the accuracy of SER in human speech noise conditions.
\item We investigate the impact of human speech noise on our framework, especially on same-gender mixture and different-gender mixture. Results indicate that our framework performs better in different-gender mixture.

\end{itemize}

\section{Proposed method}
\label{sec:method}
\subsection{Framework overview}
\label{ssec:framework}

We assume that an observed time-domain mixture signal in human speech noise conditions is defined as
\begin{align}
  \boldsymbol{y}=\boldsymbol{s}_{0}+ \sum_{i=1}^{I-1}\boldsymbol{s}_{i},
\end{align}
where \(I\) is the number of speakers in the mixture, \(\boldsymbol{s}_{i}\) for \(i = 0,...,I-1\) is the speech signal of the \(i\)th speaker. In particular, \(i = 0\) indicates the speech signal of the target speaker. Directly using the mixture signal \(\boldsymbol{y}\) corrupted by overlapping speakers for SER task, would cause the poor SER performance, as the SER model cannot selectively focus on a single speaker. Therefore, the goal of the proposed framework is to extract the target speaker signal \(\boldsymbol{s}_{0}\) from the mixture signal \(\boldsymbol{y}\) for SER training. 

\begin{figure}[t]
  \centering
  \includegraphics[width=0.47\textwidth]{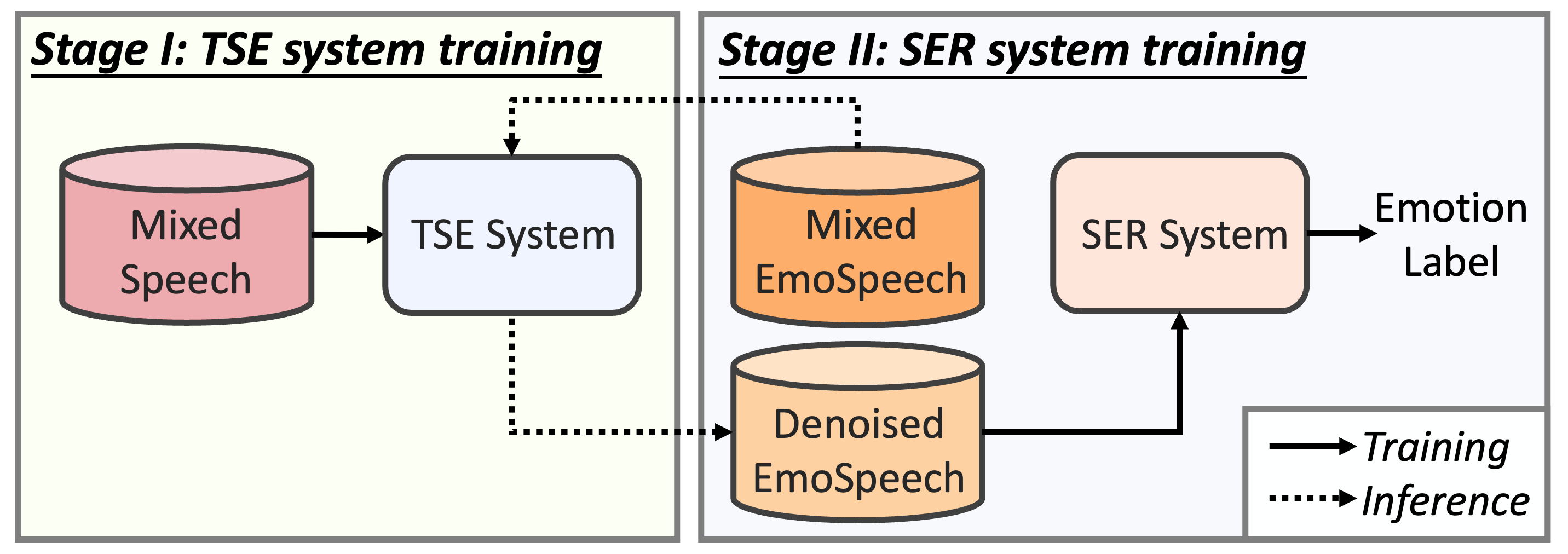}
  \caption{Two-stage framework with TSE-SER-base.} 
  \label{fig:Method1}
  % \vspace{-5pt}
  
\end{figure}

\begin{figure}[t]
  \centering
  \includegraphics[width=0.47\textwidth]{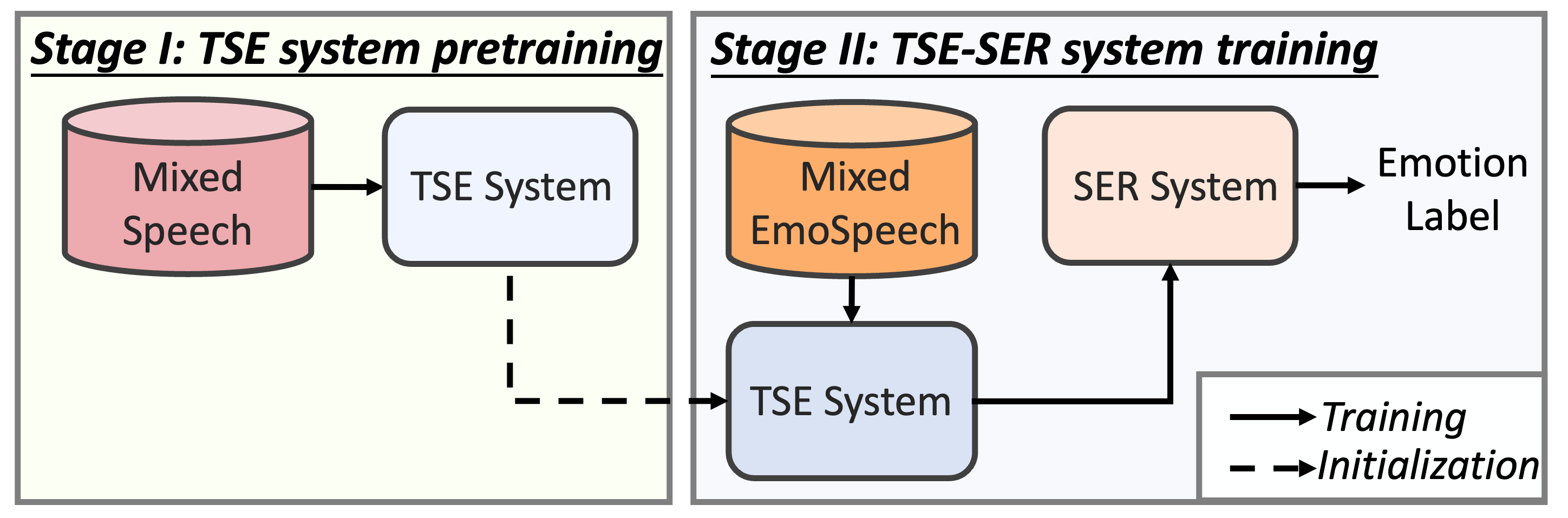}
  \caption{Two-stage framework with TSE-SER-ft.}
  \label{fig:Method2}
  % \vspace{-15pt}
\end{figure}

As illustrated in Figure~\ref{fig:Method1}, the framework contains two stages: First, the TSE model is trained using a large-scale mixed-speech corpus. This ensures a well-trained TSE model that can extract high-quality speech of the target speaker given the input speech and the enrollment utterance of that speaker. Note that we use a neutral speech utterance for enrollment, which is a more convenient and realistic setting than using corresponding emotional speech. Then, we apply the TSE model as the form of data augmentation to extract the exclusive speech information of the target speaker from a mixed emotional speech corpus. This denoised corpus is used for both training and testing in SER tasks. We denote this training method as TSE-SER-base. In addition, we further propose another training method called TSE-SER-ft shown in Figure~\ref{fig:Method2}. We start from the same TSE pretraining as the first stage. In the second stage, we introduce mixed emotional speech as input, simultaneously fine-tuning the pretrained TSE model and training the SER model. This joint training process not only refines the TSE system by adjusting its parameters but also benefits the SER training.


\subsection{Target speaker extraction model}
\label{ssec:TSE_model}
TSE refers to the task of reconstructing the speech signal of the target speaker from the mixture given auxiliary information of that speaker. This process can be formulated as
\begin{align}
  \hat{\boldsymbol{s}}_{0}=g(\boldsymbol{y}, \boldsymbol{a}_{0}),
\end{align}
where \(\hat{\boldsymbol{s}}_{0}\) is the estimated speech of the target speaker, \(\boldsymbol{a}_{0}\) is the enrollment utterance of the target speaker, \(g\) represents the transformation carried out by the TSE system.

In this work, we adopt time-domain SpeakerBeam (TD-SpeakerBeam) \cite{Vzmolikova19_Speakerbeam, Delcroix20_Speakerbeam} as the TSE model. TD-SpeakerBeam consists of an auxiliary network and a speech extraction network, represented by \(h\) and \(f\), respectively. The auxiliary network \(h\) accepts the enrollment utterance \(\boldsymbol{a}_{0}\) and computes an embedding vector, denoted by \(\boldsymbol{E}_{0}\), to represent the acoustic characteristics of the target speaker, i.e.,
\begin{align}
  {\boldsymbol{E}}_{0}=h(\boldsymbol{a}_{0}).
\end{align}
Then, the speech extraction network \(f\) accepts the mixture signal \(\boldsymbol{y}\) and the embedding vector \(\boldsymbol{E}_{0}\) of the target speaker as inputs to predict the speech signal of the target speaker, i.e., 
\begin{align}
  \hat{\boldsymbol{s}}_{0}=f(\boldsymbol{y}, \boldsymbol{E}_{0}),
\end{align}
where \(f\) comprises an encoder $\mathcal{E}$, a mask estimator $\mathcal{B}$, and a decoder $\mathcal{D}$. This process is formulated as
\setlength{\jot}{3pt} % 减少公式之间的间距,使用 gather 环境使公式居中且编号
\begin{gather}
  {\boldsymbol{Y}} =\mathcal{E}(\boldsymbol{y}), \\
  {\boldsymbol{M}_0} =\mathcal{B}(\boldsymbol{Y}, \boldsymbol{E}_0), \\
  \hat{\boldsymbol{s}}_{0} =\mathcal{D}(\boldsymbol{Y} \odot \boldsymbol{M}_0),
\end{gather}
where $\odot$ denotes element-wise multiplication. The mixture signal \(\boldsymbol{y}\) is fed into the encoder $\mathcal{E}$ that is represented by a 1D convolution layer. Then, the mask estimator $\mathcal{B}$ maps the output \(\boldsymbol{Y}\) of the encoder $\mathcal{E}$ to a mask $\boldsymbol{M}_0$ for the target speaker, utilizing multiple convolution blocks. In particular, a multiplicative adaptation layer, accepting the embedding vector \(\boldsymbol{E}_{0}\) of the target speaker as auxiliary information, is inserted between the first and second blocks to drive the network towards extracting the target speaker. Finally, the mask $\boldsymbol{M}_0$ and the encoder features \(\boldsymbol{Y}\) are fed into a 1D deconvolution layer-based decoder $\mathcal{D}$, to output the time-domain signal of the target speaker.

% \subsection{Speech emotion recognition models}
% \label{ssec:SER_model}
% We select ShiftCNN \cite{Shen23_Temporalshift} as the SER model, which has shown advanced performance in clean environments. ShiftCNN introduces temporal shifts for pretrained speech representations of SER task to address the issue of channel alignment.

\subsection{Loss functions}
\label{ssec:loss_function}
In the first stage, TSE-SER-base and TSE-SER-ft use scale-invariant source-to-noise ratio (SiSNR) \cite{Le19_SISDR} as the loss for TSE training. In the second stage, TSE-SER-base uses cross entropy (CE) loss for SER training, whereas TSE-SER-ft jointly trains the pretrained TSE model and SER model, considering both SiSNR and CE losses. The second stage losses of TSE-SER-base (${L}_{base}$) and TSE-SER-ft (${L}_{ft}$) are represented as
\begin{gather}
  {{L}_{base}} = {L}_{CE}, \\
  {{L}_{ft}} = {L}_{SiSNR} + {L}_{CE}.
\end{gather}

\section{Experimental evaluations}
\subsection{Datasets}
\label{ssec:Datasets}
In this work, we designed two kinds of datasets for comparable experiments: 1) the clean emotional dataset, and 2) the emotional dataset mixed with human speech noise. For the latter, two different human speech datasets were used as noise. All the mentioned datasets were sampled at 16 kHz. 
\label{section:dataset}
{\par\noindent\textbf{IEMOCAP:} The Interactive Emotional Dyadic Motion Capture (IEMOCAP) corpus \cite{Busso08_IEMOCAP}, consisting of approximately 12 hours of recordings, includes five dyadic sessions, each with one English male speaker and one English female speaker. For our experiments, we used IEMOCAP as the clean emotional dataset, where we considered only four emotional categories of happy, angry, sad, and neutral. Note that “excited” was merged with “happy” to ensure category balance \cite{Chen23_Exploring, Zen22_CoAttention, guo2021representation, shi2023emotion, shi2024Multimodal}.}

 {\par\noindent\textbf{LibriSpeech:} The LibriSpeech corpus \cite{Panayotov15_Librispeech} contains about 1000 hours of read English speech. For our experiments, 105 hours of this corpus were chosen as a source of human speech noise.}

{\par\noindent\textbf{ESD:} The Emotional Speech Database (ESD) corpus  \cite{Zhou22_ESD} comprises about 29 hours of recordings from 10 English speakers and 10 Chinese speakers. For our experiments, we considered only the English part as another source of human speech noise.}

In order to more accurately evaluate the performance of the proposed framework, we adopted leave-one-session-out 5-fold cross-validation to test all the models. Note that we used the following terms to represent different datasets designed: \textit{Clean} means a clean set, \textit{Noisy} means a dataset mixed with human speech noise, and \textit{Denoised} indicates a dataset denoised from \textit{Noisy} by the pretrained TSE model. 

% 介绍每个实验的具体设置，包括干净训练集和测试集，具体的混合方式，TSE预训练的训练集，数据集的表示方式
\subsection{Experimental procedure}
\label{ssec:procedure_experiment}
The first experiment investigated the impact of human speech noise on SER (see Section~\ref{sssec:result1}). The noisy dataset was generated by randomly selecting utterances from different speakers in LibriSpeech and mixing them with clean data of IEMOCAP. The number of speakers in the mixture of speech was limited to two.

The second and third experiments explored the effect of TSE method on SER and the proposed training methods on our framework (see Sections~\ref{sssec:result2} and~\ref{sssec:result3}). We used the same dataset as the first experiment for the second stage of TSE-SER-base and TSE-SER-ft. We adopted LibriMix \cite{Cosentino20_librimix} where 100 hours of LibriSpeech were additionally used to generate mixtures for TSE pretraining. Furthermore, for the target speaker in the mixture, we randomly chose one neutral utterance of this speaker that does not belong to the mixture as the enrollment utterance.

The fourth experiment explored the impact of gender states of mixtures on SER (see Section~\ref{sssec:result4}). We used the same clean speech from the first experiment and used ESD as noise. We generated two types of mixtures: same-gender mixture and different-gender mixture, where the first type was stipulated that two speakers have the same genders, while the latter type required two speakers of opposite genders.

\subsection{Implementation and metrics}
\label{ssec:implementation_metrics}
To build TSE model, we followed an open-source SpeakerBeam implementation$\footnote{\url{https://github.com/BUTSpeechFIT/speakerbeam}}$. For SER model, we used ShiftCNN \cite{Shen23_Temporalshift}, which has shown advanced performance in clean environments, adopting the same hyperparameters as \cite{Shen23_Temporalshift}. All implemented models in experiments are shown in Table~\ref{tab:systems_name}. 

For evaluation metrics of SER, we used the unweighted accuracy (UA) and weighted accuracy (WA). UA was the mean of the accuracies for each individual class while WA represented the ratio of correctly predicted samples to the total number of samples. In addition, we used scale-invariant signal-to-distortion ratio (SI-SDR) and scale-invariant signal-to-distortion ratio improvement (SI-SDRi) to evaluate the performance of TSE model.
% TABLE1: SYSTEM NAME
\begin{table}[t]\small
  \caption{Definition of the models used in experiments.}
  \label{tab:systems_name}
  \centering
  \begin{tabular}{ccc}
    \toprule[1pt]
    \textbf{\begin{tabular}[c]{@{}c@{}}Short\\ Name\end{tabular}}    & \textbf{Model}                       & \textbf{Method} \\
    \midrule
    SB                      & TD-SpeakerBeam \cite{Delcroix20_Speakerbeam} & -                 \\
    SC                & ShiftCNN \cite{Shen23_Temporalshift} & -                 \\
    SB\_SC            & TD-SpeakerBeam + ShiftCNN               & TSE-SER-base                \\
    SB+SC             & TD-SpeakerBeam + ShiftCNN               & TSE-SER-ft               \\
    \bottomrule[1pt]
  \end{tabular}
  \vspace{-5pt}
\end{table}

\subsection{Experimental results}
\label{ssec:experimental_results}
To conduct a comparative study of all the experiments, aside from the proposed TSE-SER-base and TSE-SER-ft, we built two typical SER baselines using ShiftCNN, which were directly trained on \textit{Clean} and \textit{Noisy}, referred to as clean SER model and noisy SER model, respectively.

\subsubsection{The impact of human speech noise on SER}
\label{sssec:result1}
To clarify the impact of human speech noise on SER, we compare the clean SER model and the noisy SER model. As shown in Table~\ref{tab:clean_noise}, the clean SER model obtains an accuracy of over 70\% for both UA and WA in the clean test set. But their performance drops significantly on the noisy test data, with a maximum decrease of up to 23.09\% and 25.62\% in terms of WA and UA. These results demonstrate that the clean SER model is fragile against human speech noise. Furthermore, when the SER model uses noisy data for training, the relatively better performance can be observed. Nonetheless, the results are still significantly worse than those of the clean SER model on the clean test set, suggesting deficient robustness. We argue that human speech noise severely hinders the SER model from establishing an effective mapping to the target emotional speech with the direct training.

\subsubsection{The effect of TSE method on SER}
\label{sssec:result2}
To verify the effectiveness of the proposed framework on SER, we compare the performance of the SER model using the TSE method with those of the SER baselines. As presented in Table~\ref{tab:noise_denoise}, we first observe that the results of the clean SER model on denoised test set are not ideal, demonstrating that the clean model is unable to adapt to the denoised speech with distorted properties. We hence design the corresponding system trained on denoised data using TSE-SER-base, referred to as SB\_SC in Table~\ref{tab:noise_denoise}. Our proposed system SB\_SC performs significantly better than the other systems. Especially, the UA reaches 63.42\%, which is a 9.48\% increase compared to the noisy SER model in noisy conditions. Meanwhile, for the unbalanced training data from IEMOCAP corpus, the WA result is also competitive, closely resembling the UA results. The overall results demonstrate that the proposed framework with TSE-SER-base adapts well to human speech noise, showcasing effectiveness and robustness.

% TABLE2:
\begin{table}[t]\small
\caption{Performance of SER in clean and human speech noise conditions.}
\label{tab:clean_noise}
\centering
\begin{tabular}{ccccc}
\toprule[1pt]
\textbf{Model} & \textbf{Train Set} & \textbf{Test Set}  & \textbf{\begin{tabular}[c]{@{}c@{}}WA \\ (\%)\end{tabular}} & \textbf{\begin{tabular}[c]{@{}c@{}}UA \\ (\%)\end{tabular}} \\ \midrule
\multirow{3}{*}{SC} & Clean      & Clean       & \textbf{70.20}   & \textbf{71.64} \\
                    & Clean      & Noisy       & 47.11 & 46.02        \\ 
                    & Noisy      & Noisy       & 53.71 & 53.94        \\
\bottomrule[1pt]
\end{tabular}
% \vspace{-5pt}
\end{table}

% TABLE3:
\begin{table}[t]\small
\caption{Comparison with TSE-SER-base and the SER baselines.}
\label{tab:noise_denoise}
\centering
\begin{tabular}{ccccc}
\toprule[1pt]
\textbf{Model} & \textbf{Train Set} & \textbf{Test Set}  & \textbf{\begin{tabular}[c]{@{}c@{}}WA \\ (\%)\end{tabular}} & \textbf{\begin{tabular}[c]{@{}c@{}}UA \\ (\%)\end{tabular}} \\ \midrule
\multirow{2}{*}{SC} & Noisy      & Noisy        & 53.71 & 53.94        \\
                    & Clean      & Denoised     & 51.12 & 49.50        \\ 
\midrule
SB\_SC              & Denoised   & Denoised     & \textbf{62.20} & \textbf{63.42}        \\
\bottomrule[1pt]
\end{tabular}
% \vspace{-5pt}
\end{table}

% TABLE4:
\begin{table}[t]\small
\caption{Comparison with TSE-SER-base and TSE-SER-ft.}
\label{tab:train_method}
\centering
\begin{tabular}{cccc}
\toprule[1pt]
\textbf{Model} & \textbf{Method}  & \textbf{\begin{tabular}[c]{@{}c@{}}WA \\ (\%)\end{tabular}} & \textbf{\begin{tabular}[c]{@{}c@{}}UA \\ (\%)\end{tabular}} \\ \midrule
SB\_SC & TSE-SER-base         & 62.20  & 63.42        \\
SB+SC  & TSE-SER-ft           & \textbf{67.02} & \textbf{68.27}        \\ 
\bottomrule[1pt]
\end{tabular}
% \vspace{-5pt}
\end{table}

% \vspace{-4pt}
% TABLE5:
\begin{table}[t]\small
\caption{Comparison with TSE-SER-base and the SER baselines on same- and different-gender mixtures.}
\label{tab:different_genders}
\centering
\begin{tabular}{cccccc}
\toprule[1pt]
\textbf{Model} & \textbf{Train Set} & \textbf{Test Set} & \textbf{\begin{tabular}[c]{@{}c@{}}Gender \\ State\end{tabular}} & \textbf{\begin{tabular}[c]{@{}c@{}}WA \\ (\%)\end{tabular}} & \textbf{\begin{tabular}[c]{@{}c@{}}UA \\ (\%)\end{tabular}} \\ \midrule
\multirow{3}{*}{SC} & Clean         & Noisy      & \multirow{3}{*}{Same}      & 48.00        & 46.50        \\
                    & Noisy      & Noisy      &        & 54.67        & 54.84        \\ 
                    & Clean         & Denoised &        & 45.09        & 43.43        \\ \midrule
SB\_SC              & Denoised & Denoised & Same      & \textbf{55.09}        & \textbf{55.95}        \\ \midrule
\multirow{3}{*}{SC} & Clean         & Noisy      & \multirow{3}{*}{Different} & 46.62        & 44.60        \\
                    & Noisy     & Noisy     &           & 54.29        & 54.20        \\
                    & Clean         & Denoised &           & 48.87        & 47.35        \\ \midrule
SB\_SC              & Denoised & Denoised & Different & \textbf{59.75}        & \textbf{61.32}        \\
\bottomrule[1pt]
\end{tabular}
% \vspace{-5pt}
\end{table}

% TABLE6
% \vspace{-20pt}
\begin{table}[t]\small
% \vspace{5pt}
\caption{Ablation study for TSE-SER-base. “Same” and "Different" indicate gender states.}
\label{tab:ablation}
\centering
\begin{tabular}{ccccc}
\toprule[1pt]
\textbf{Model} &
  \textbf{Train set} &
  \textbf{Test set} &
  \textbf{\begin{tabular}[c]{@{}c@{}}WA \\ (\%)\end{tabular}} &
  \textbf{\begin{tabular}[c]{@{}c@{}}UA \\ (\%)\end{tabular}} \\ \midrule
\multirow{4}{*}{\begin{tabular}[c]{@{}c@{}c@{}c@{}c@{}}\\ SB\_SC\end{tabular}} &
  \multirow{2}{*}{\begin{tabular}[c]{@{}c@{}c@{}c@{}c@{}}\vspace{-1pt} Denoised \\ (Same)\end{tabular}} &
  \begin{tabular}[c]{@{}c@{}c@{}c@{}c@{}} Denoised (Same) \end{tabular} &
  55.09 &
  55.95 \\  
 &  & \begin{tabular}[c]{@{}c@{}}Denoised (Different)\end{tabular} & 59.12 & 60.46 \\ \addlinespace[1pt] \cline{2-5} \addlinespace[1.5pt]
 &
  \multirow{2}{*}{\begin{tabular}[c]{@{}c@{}c@{}c@{}c@{}}\vspace{-1pt} Denoised \\ (Different)\end{tabular}} &
  \begin{tabular}[c]{@{}c@{}}Denoised (Same)\end{tabular} &
  55.32 &
  56.40 \\ 
 &  & \begin{tabular}[c]{@{}c@{}}Denoised (Different)\end{tabular} & 59.75 & 61.32 \\ 
\bottomrule[1pt]
\end{tabular}
\end{table}

\subsubsection{The effect of training methods on TSE--SER framework}
\label{sssec:result3}
We compare the performance of TSE-SER-base and TSE-SER-ft. As indicated in Table~\ref{tab:train_method}, TSE-SER-ft significantly outperforms TSE-SER-base. Moreover, the UA of TSE-SER-ft reaches a 14.33\% improvement compared with the noisy SER model presented in Table~\ref{tab:noise_denoise}. The SI-SDR of the noisy speech before being processed by the TSE model is 0.09 dB. We calculate the SI-SDRi for the corresponding TSE models of TSE-SER-base and TSE-SER-ft to be 7.68 dB and 12.90 dB, respectively, verifying that TSE-SER-ft can improve TSE performance. Therefore, the TSE fine-tuning, applied to the noisy dataset containing the task-specific emotional data, enables the TSE model to extract purer emotional-related acoustic features to benefit the SER training.

\subsubsection{The impact of gender states of mixtures on SER}
\label{sssec:result4}
Table~\ref{tab:different_genders} shows the results of TSE-SER-base, the SER baselines on same- and different-gender mixtures. First, the clean SER models unsurprisingly gain the lowest performance. Moreover, the noisy SER model shows a non-obvious trend on the noisy test set across both gender states, whereas the clean SER model in same-gender mixture clearly outperforms that in different-gender mixture, demonstrating better adaptability of SER models to same-gender mixture. In addition, besides the results of TSE-SER-base being all better than those of other systems for both same- and different-gender mixtures, we can see a significant performance gap of TSE-SER-base for dealing with same- and different-gender mixtures, which is opposite to the finding for the clean SER model in noisy conditions. Since our framework uses the TSE model, we conjecture that same-gender mixture with the similar acoustic characteristics is more difficult for TSE model to separate. Before being processed by the TSE model, the SI-SDR of same- and different-gender mixtures are 0 dB and 0.02 dB. We calculate the SI-SDRi for TSE model to be 1.09 dB and 5.22 dB for same- and different-gender mixtures, respectively. This also explains why the clean SER model on the denoised test set gives better results in different-gender mixture.

We further conduct an ablation study for TSE-SER-base using two SB\_SC systems from Table~\ref{tab:different_genders} on test sets denoised from the same- and different-gender mixtures. We have an interesting finding in Table~\ref{tab:ablation} that using training data denoised from different-gender mixture can enhance our model performance on both test sets denoised from same- and different-gender mixtures. We argue that the pretrained TSE model can extract higher-quality denoised data from different-gender mixture, thus finalizing a higher-performance SER model.

\section{Conclusion}
In this work, we presented a two-stage framework to mitigate the impact of human speech noise on SER. Based on the framework, we designed two training methods, TSE-SER-base and TSE-SER-ft. The effectiveness and robustness of both methods have been verified. Moreover, we investigated the impact of human speech noise on SER, especially on same- and different-gender mixtures. In the future, we plan to explore how our proposed framework performs on human speech noise with different attributes, such as emotion classes and languages. Another possible direction involves multi-interference environments, such as noisy and reverberant environments.

\section*{Acknowledgment}
This work was partly supported by JST CREST Grant Number JPMJCR19A3, Japan, and JSPS KAKENHI Grant Number 21H05054. In addition, this work was also financially supported by JST SPRING, Grant Number JPMJSP2125. The author would like to take this opportunity to thank the “THERS Make New Standards Program for the Next Generation Researchers.”

\printbibliography

\end{document}